# Transverse thermoelectric generation using magnetic materials

Ken-ichi Uchida[1-3,a)], Weinan Zhou[1], and Yuya Sakuraba[1,4,a)]

**AFFILIATIONS**

[1] National Institute for Materials Science, Tsukuba 305-0047, Japan

[2] Institute for Materials Research, Tohoku University, Sendai 980-8577, Japan

[3] Center for Spintronics Research Network, Tohoku University, Sendai 980-8577, Japan

[4] PRESTO, Japan Science and Technology Agency, Saitama 332-0012, Japan

[a)] Authors to whom correspondence should be addressed: UCHIDA.Kenichi@nims.go.jp and SAKURABA.Yuya@nims.go.jp

**ABSTRACT**

The transverse thermoelectric effect refers to the conversion of a temperature gradient into a transverse charge current, or vice versa, which appears in a conductor under a magnetic field or in a magnetic material with spontaneous magnetization. Among such phenomena, the anomalous Nernst effect in magnetic materials has been receiving increased attention from the viewpoints of fundamental physics and thermoelectric applications owing to the rapid development of spin caloritronics and topological materials science. In this research trend, a conceptually different transverse thermoelectric conversion phenomenon appearing in thermoelectric/magnetic hybrid materials has been demonstrated, enabling the generation of a large transverse thermopower. Here, we review the recent progress in fundamental and applied studies on the transverse thermoelectric generation using magnetic materials. We anticipate that this perspective will further stimulate research activities on the transverse thermoelectric generation and lead to the development of next-generation thermal energy harvesting and heat-flux sensing technologies.

Thermoelectric generation technologies have been considered as future independent power sources for Internet of Things applications because electricity can be generated from waste heat.[1-3] Existing thermoelectric devices are based on the Seebeck effect owing to its relatively high thermoelectric conversion efficiency. However, because the Seebeck effect is the longitudinal thermoelectric effect in which a charge current is generated in the direction parallel to the applied temperature gradient [Fig. 1(a)], the thermoelectric modules have complicated structures. As shown in Fig. 1(b), to increase the thermoelectric output, many pairs of two different thermoelectric materials must be connected in series and the materials must be lengthened in the direction of the temperature gradient. This complicated structure limits the durability and flexibility of thermoelectric devices and hinders their wider applications.

One approach to overcome this problem is to use transverse thermoelectric effects. As they generate a charge current in the direction perpendicular to a temperature gradient, the thermoelectric generation is possible simply by forming a material onto a heat source surface. By utilizing this feature, the output voltage (power) can be enhanced by elongating the device length (enlarging the device area) perpendicular to the temperature gradient without constructing three-dimensional serial junctions. The thermoelectric modules based on transverse thermoelectric effects have a simple structure in which conductors are connected in series along the heat source surface, enabling the efficient use of waste heat, reducing the cost of modules, and improving their durability and flexibility. Thus, transverse thermoelectric effects have the potential to solve the problems associated with conventional Seebeck devices.

One of the fundamental transverse thermoelectric effects is the Nernst effect. The ordinary Nernst effect (ONE), discovered by W. Nernst and A. V. Ettingshausen in the late 19th century,[4] refers to the generation of a charge current in the direction perpendicular to both a temperature gradient and external magnetic field applied to a conductor. Although ONE in semiconductors and semimetals exhibit a substantially large transverse thermopower, it has not been applied in practice because its operation requires large magnetic fields.[5-8] In contrast, in magnetic materials with spontaneous magnetization, the Nernst effect appears even in the absence of magnetic fields owing to the spin-orbit interaction, which is called the anomalous Nernst effect (ANE)[9-55] [Fig. 1(c)]. Therefore, the transverse thermopower $S$ in a magnetic material under a magnetic field is described as the summation of the contributions proportional to the magnitude of the magnetic field $H$ and magnetization $M$:

$$S = Q_H \mu_0 H + Q_M \mu_0 M, \qquad (1)$$

where $\mu_0$ is the vacuum permeability and $Q_{H(M)}$ is the proportionality factor of each term.

Equation (1) shows that the anomalous Nernst coefficient $S_{ANE}$ (= $Q_M\mu_0 M_s$ with $M_s$ being the saturation magnetization) can be extracted by extrapolating the $H$ dependence of $S$ from the high field region, in which $M$ is saturated, to zero field. When the total transverse thermopower is dominated by ANE, the $H$ dependence of $S$ follows the $M$-$H$ curve of the magnetic material. Based on Eq. (1), $S_{ANE}$ is often compared in terms of $M$ to discuss the scaling behavior.[25,36] However, since ANE and related transport properties are estimated for uniformly magnetized materials, it is natural to compare the coefficients in terms of $M_s$, not $M$. Importantly, as shown in Fig. 2(a), $S_{ANE}$ is not correlated with $M_s$ in many materials including Fe, Ni, and simple binary alloys.

The charge current density $\mathbf{j}_{c,ANE}$ driven by ANE is expressed as

$$\mathbf{j}_{c,ANE} = \sigma S_{ANE}(\mathbf{m} \times \nabla T), \qquad (2)$$

where $\sigma$ is the longitudinal electrical conductivity, $\mathbf{m}$ is the unit vector of magnetization, and $\nabla T$ is the temperature gradient applied to the magnetic material. When $\mathbf{m}$ aligns perpendicular to $\nabla T$, ANE works as a thermoelectric generator. In an open circuit condition with $\mathbf{m}$ and $\nabla T$ respectively being along the $z$ and $x$ directions, the following equation holds for the $y$ direction: $\mathbf{j}_{c,ANE} + \sigma \mathbf{E} = 0$, where $\mathbf{E}$ is the electric field appearing due to the charge accumulation induced by $\mathbf{j}_{c,ANE}$. Thus, one obtains

$$\mathbf{E} = S_{ANE}(\nabla T \times \mathbf{m}), \qquad (3)$$

and the transverse thermoelectric voltage $V_y = E_y L_y$ becomes observable in ANE experiments, where $E_y$ and $L_y$ are the magnitude of $\mathbf{E}$ and the length of the material along the $y$ direction, respectively. Here, the $\mathbf{E}$ direction is opposite to the direction of the ANE-driven electric field $\mathbf{E}_{ANE}$ (= $\mathbf{j}_{c,ANE}/\sigma$), which may confuse the definition of $S_{ANE}$, although Eqs. (2) and (3) are based on the same argument (note also that the definition, or sign, of $S_{ANE}$ is sometimes different in different papers). Equation (3) indicates that the thermoelectric output of ANE can be actively controlled through the manipulation of $\mathbf{m}$. Owing to this feature and the aforementioned simple device structure, ANE is expected to realize simple and versatile thermal energy harvesting or heat-flux sensing applications [Fig. 1(d)].

Although primary studies on ANE were conducted a long time ago and research activities were quite limited, ANE has received renewed attention in the field of spin caloritronics.[56,57] In the early days of spin caloritronics, many experiments were performed to distinguish ANE from the thermo-spin conversion called the spin Seebeck effect because these phenomena exhibit a similar symmetry to each other.[58-62] Through such activities, spin caloritronics has dramatically promoted ANE studies from the viewpoints of both fundamental physics and thermoelectric applications. In particular, since the observation of giant ANE in magnetic topological materials in 2018,[31] physics and materials science studies have further accelerated and become a trend in condensed matter physics. Large ANE has also been observed in various materials, including Heusler compounds,[31,32,36,44,47,51,53] ferromagnetic binary alloys,[20,40,46,48,49] and multilayer films,[21,55] of which $S_{ANE}$ is an order of magnitude larger than that of Fe.[12,40,46] In parallel with the works on ANE, the anomalous Ettingshausen effect, the Onsager reciprocal of ANE, has also been observed in various magnetic materials in both bulk and film forms by means of active thermal detection techniques.[63-74] Such measurements reveal that rare-earth permanent magnets can be promising candidate materials for transverse thermoelectric conversion.[70,74] More recently, not only plain magnetic materials but also hybrid structures comprising thermoelectric and magnetic materials have begun to be used, and a conceptually different effect called the Seebeck-driven transverse thermoelectric generation (STTG) has been demonstrated.[75-77] In light of these research activities, in this perspective, we review recent studies on transverse thermoelectric generation using magnetic materials and discuss its potential applications. Here, we focus on ANE and STTG; although the spin Seebeck effect also enables transverse thermoelectric generation, it is not covered in this perspective.

To realize practical applications of ANE, it is necessary to find and develop magnetic materials with large $S_{ANE}$. Design guidelines for materials suitable for ANE can be obtained by separating $S_{ANE}$ into two components:

$$S_{ANE} = \rho_{xx}\alpha_{xy} - \rho_{AHE}\alpha_{xx}, \qquad (4)$$

where $\rho_{xx}$, $\rho_{AHE}$, and $\alpha_{xx}$ ($\alpha_{xy}$) are the longitudinal electrical resistivity, anomalous Hall resistivity, and diagonal (off-diagonal) component of the thermoelectric conductivity tensor, respectively. The second term on the right-hand side of Eq. (4), $\rho_{AHE}\alpha_{xx}$, appears as a consequence of the fact that the longitudinal carrier flow induced by the Seebeck effect is bent due to the anomalous Hall effect (AHE).[78] The first term, $\rho_{xx}\alpha_{xy}$, is usually regarded as intrinsic ANE because $\alpha_{xy}$ directly converts $\nabla T$ into a transverse electric field. A recent trend in improving $S_{ANE}$ is to find materials with large $\alpha_{xy}$, in which the Berry curvature of the electronic bands near the Fermi level plays an important role. Materials with topological band structures show large $\alpha_{xy}$ values due to the Berry curvature. The resultant $S_{ANE}$ of 6-8 $\mu$VK$^{-1}$ in Co-based Heusler compounds is the current record high at room temperature.[31,36,47,51,53] The large $S_{ANE}$ in SmCo$_5$-type magnets is also believed to be dominated by large $\alpha_{xy}$ due to the intrinsic mechanism.[70,74] Importantly, when ANE originates from electronic band structures, $S_{ANE}$ and $\alpha_{xy}$ have no correlation with the saturation magnetization (Fig. 2). Another strategy for enhancing ANE is to optimize magnetic multilayer structures; ANE in alternately stacked ferromagnetic metal/nonmagnetic metal multilayers is enhanced by increasing the number of interfaces per unit volume.[21,55] Although the enhancement of ANE in multilayers seems to be a universal behavior, the microscopic mechanism responsible for this phenomenon has not been clarified so far. As shown here, both the bulk transport properties and interface engineering are important for obtaining large ANE. Although ANE studies have progressed rapidly in recent years, the obtained $S_{\text{ANE}}$ values are still smaller than 10 $\mu$VK$^{-1}$, which is 1-2 orders of magnitude smaller than the Seebeck coefficients of thermoelectric materials in practical use. Therefore, further breakthrough developments in physics and materials science are needed for the applications of transverse thermoelectric generation. As a part of such efforts, STTG was proposed, which will be discussed later.

In general, ANE can be characterized by measuring the magnetic field dependence of a transverse thermopower in magnetic materials. Following Eq. (3), the ANE-induced thermopower exhibits the odd dependence on the **m** direction. In contrast to isotropic bulk materials, ANE measurements for thin films are performed in two different configurations because of the huge difference between the in-plane and out-of-plane dimensions.[60-62] One configuration is the in-plane magnetized (IM) configuration, where the magnetic field **H** ($\nabla T$) is applied along the in-plane (out-of-plane) direction [inset of Fig. 3(a)]. The other configuration is the perpendicularly magnetized (PM) configuration, where **H** ($\nabla T$) is applied along the out-of-plane (in-plane) direction [inset of Fig. 3(b)]. Figures 3(a) and 3(b) show an example of the experimental results of ANE measured in the IM and PM configurations, respectively, for the two 50-nm-thick Co$_2$MnGa thin films with different composition ratios: Co$_{53.0}$Mn$_{23.8}$Ga$_{23.2}$ (referred to as Co$_2$MnGa-A) and Co$_{41.4}$Mn$_{27.9}$Ga$_{30.7}$ (referred to as Co$_2$MnGa-B). The Co$_2$MnGa-A and Co$_2$MnGa-B films were epitaxially grown on single-crystalline MgO (001) substrates. Because of the strong demagnetization field in the out-of-plane direction, the remanent magnetization is stabilized in the in-plane direction, resulting in large ANE voltage even in the absence of an external magnetic field in the IM configuration [Fig. 3(a)], whereas no ANE voltage is generated at zero field in the PM configuration [Fig. 3(b)]. As depicted in Fig. 1(d), the IM configuration is suitable for thermal energy harvesting and heat flux sensing because ANE-based thermoelectric generation is realized simply by forming films onto heat sources. Large ANE voltage owing to the remanent magnetization in the IM configuration is also preferable for practical applications. In contrast, in the IM configuration, it is difficult to estimate $S_{\text{ANE}}$ quantitatively because the temperature difference between the top and bottom of the thin films has to be quantified. Therefore, for the quantitative estimation of $S_{ANE}$, the PM configuration is widely used, although it often requires a large magnetic field to align **m** of films along the out-of-plane direction to overcome the strong demagnetization field. In the PM configuration, the magnitude and distribution of $\nabla T$ can be exactly measured by several experimental techniques, such as thermometers attached on hot and cold sides of the substrate or heat baths,[20,21,34,54] on-chip thin-film thermometers grown on the substrate,[32,44,48] and an infrared camera.[40,46,47,53,55,65] Thus, $E_y$ can be normalized by the temperature gradient $\nabla_x T$ along the x direction, enabling the estimation of $S_{ANE}$ by taking the zero-field intercept of the magnetic field dependence of $E_y/\nabla_x T$ [see Eq. (1)].

In Figs. 3(a) and 3(b), one can see the large difference in the ANE-induced thermopower between the Co$_2$MnGa-A and Co$_2$MnGa-B films, although the X-ray diffraction (XRD) patterns for these films show almost identical crystal structure and $L2_1$-type atomic ordering [Fig. 3(c)]. One of the reasons for such a dramatic variation of ANE in almost the same materials is the large difference in the first term of Eq. (4), $\rho_{xx}\alpha_{xy}$, through the intrinsic $\alpha_{xy}$. To obtain the intrinsic $\alpha_{xy}$ from the anomalous Hall conductivity $\sigma_{xy}$ ($= \rho_{AHE}/(\rho_{xx}^2 + \rho_{AHE}^2)$) originating from the Berry curvature, the following expression is often used in first-principles calculations:

$$\alpha_{xy} = -\frac{1}{eT}\int d\varepsilon \left(-\frac{\partial f}{\partial \varepsilon}\right)(\varepsilon - \mu)\sigma_{xy}(\varepsilon) \qquad (5)$$

where $f = 1/\left[\exp((\varepsilon-\mu)/k_B T)+1\right]$ is the Fermi distribution function with $-e$ ($e > 0$), $\varepsilon$, $\mu$, and $k_B$ respectively being the electron charge, energy, chemical potential, and Boltzmann constant. Figure 3(e) shows the $\varepsilon - \varepsilon_F$ dependence of $\alpha_{xy}$ calculated for stoichiometric Co$_2$MnGa using Eq. (5) based on $\sigma_{xy}$ shown in Fig. 3(d), where $\varepsilon_F$ is the Fermi energy. The $\alpha_{xy}$ value shows a steep change around the Fermi level, $\varepsilon = \varepsilon_F$, because of the peak of $\sigma_{xy}$ near the Fermi level. This calculation suggests that the large difference in the ANE-induced thermopower between the Co$_2$MnGa-A and Co$_2$MnGa-B films is mainly attributed to a different position of the Fermi level caused by their composition difference, which was directly proven by photoemission spectroscopy in a previous study.[53] Namely, the Co$_2$MnGa-A film has the Co-rich composition giving an additional valence electron compared to the stoichiometric case, which shifts the Fermi level upward by approximately 0.07 eV corresponding the peak position of the theoretical $\alpha_{xy}$. Therefore, large $\alpha_{xy}$ of 3.3 Am$^{-1}$K$^{-1}$ was experimentally obtained in the Co$_2$MnGa-A film, which is close to the theoretical $\alpha_{xy}$ of 4.2 Am$^{-1}$K$^{-1}$ at $\varepsilon - \varepsilon_F = 0.07$ eV, whereas the Co$_2$MnGa-B film exhibits much smaller $\alpha_{xy}$ of 0.8 Am$^{-1}$K$^{-1}$ because of its lower valence electron number caused by the Co-poor composition ratio. As indicated by this result, it is important to tune the Fermi level to obtain the theoretically predicted highest $\alpha_{xy}$ and the resultant large $S_{ANE}$ in various materials. It is worth mentioning here that the contribution of extrinsic mechanisms, such as skew scattering[79] and side jump,[80] on ANE has been disregarded in the previous studies and thus not been clarified so far, which might cause a disagreement between the experimental and theoretically calculated intrinsic values in the magnitude and sign of $\alpha_{xy}$. For example, as shown in Fig. 3(e), the sign of $\alpha_{xy}$ in $Co_2MnGa$ becomes negative by slightly reducing the valence electron number. In the experiment, however, positive $\alpha_{xy}$ was always obtained in the $Co_2MnGa$-B film and other $Co_2MnGa$ films having a valence electron number lower than the stoichiometry.[53] Weischenberg et al. calculated both the intrinsic and side-jump contributions on ANE in bcc Fe, hcp Co, fcc Ni, and $L1_0$-FePt and claimed their equal importance.[17] Thus, the elucidation of extrinsic contributions to ANE in various materials is important to explore materials with large $S_{ANE}$.

Toward the thermoelectric applications based on ANE, not only physics and materials science studies but also device engineering is indispensable. As described above, one of the advantages of ANE-based thermoelectric applications is the simple device structure, in which the thermoelectric voltage is easily enlarged by elongating the total length of a magnetic wire grown/attached on a heat source surface. Here, we present an experimental demonstration of large thermoelectric voltage generation in the ANE-based module in the IM configuration [Fig. 1(d)]. Figure 4(a) shows the magnetic field dependence of $V_y$ for a lateral $Co_2MnGa$-Au thermopile device, in which 50 $Co_2MnGa$ wires with a length of 10 mm, width of 50 μm, and thickness of 1 μm are connected in series through Au wires with a length of 10 mm, width of 50 μm, and thickness of 400 nm. Thus, the total length of the $Co_2MnGa$ wires reaches 500 mm. The wires were fabricated on an area of 10 × 10 mm$^2$. It is clearly found that the $V_y$ signal in the $Co_2MnGa$-Au thermopile is two orders of magnitude larger than that in the non-patterned $Co_2MnGa$ films shown in Fig. 3(a), which confirms the usefulness of the simple lateral thermopile device for ANE. A promising application of such ANE-based thermopiles is a heat flux sensor with high flexibility and low thermal resistance.[46] As shown in Fig. 4(b), the ANE signal linearly increases with the heat flux density $j_q$ flowing across the thermopile device, which works as a heat flux sensor. The slope of the $j_q$ dependence of $V_y$, which gives a sensitivity of $j_q$, was estimated to be 0.110 μVW$^{-1}$m$^2$ at $\mu_0 H$ = 0.10 T, where the magnetization of the $Co_2MnGa$ wires aligns along the field direction. This sensitivity is an order of magnitude larger than that in the prototypical $Fe_{81}Al_{19}$/Au thermopiles, which mainly originates from the longer total length of the magnetic wire and larger $S_{\text{ANE}}$ in Co$_2$MnGa than in Fe$_{81}$Al$_{19}$.[46] However, it is important to mention that the sensitivity at zero field reduces to 0.016 $\mu$VW$^{-1}$m$^2$ because of the small remanent magnetization of the narrow Co$_2$MnGa wires, which is caused by an increase in the demagnetization field and possible magnetic domain formation. Therefore, to preserve large transverse thermoelectric voltage at zero field in elongated narrow magnetic wires, magnetic materials with small magnetization and/or large uniaxial magnetic anisotropy in the in-plane direction are preferable. For a practical usage of ANE-based heat flux sensors, the sensitivity of >1 $\mu$VW$^{-1}$m$^2$ is desired. Although this sensitivity value is comparable to or smaller than that of conventional Seebeck-based heat flux sensors, ANE has strong advantages in flexibility and low thermal resistance, extending applications of heat flux sensors.

Up to now, we have outlined the recent activities on ANE. In 2021, we proposed and demonstrated the transverse thermoelectric generation different from ANE. This is named STTG.[75] Although ANE appears in a plain magnetic material, STTG appears in hybrid structures comprising thermoelectric and magnetic materials and originates from the combination of the Seebeck effect in the former and AHE in the latter, which is inspired by the second term on the right-hand side of Eq. (4). Figure 5(a) shows a schematic of the typical structure used to demonstrate STTG. When $\nabla T$ is applied to a closed circuit comprising thermoelectric and magnetic materials, a charge current is induced by the difference in the Seebeck coefficients of the materials. This charge current is in turn converted into a transverse electric field owing to AHE in the magnetic material. Based on our phenomenological calculation, the transverse thermopower in the hybrid structure shown in Fig. 5(a) is expressed as follows:[75,77]

$$S_{\text{tot}} = S_{\text{ANE}} - \frac{\rho_{\text{AHE}}}{\rho_{\text{TE}}/r + \rho_{\text{M}}} (S_{\text{TE}} - S_{\text{M}}), \tag{6}$$

where $\rho_{\text{TE(M)}}$ and $S_{\text{TE(M)}}$ are the longitudinal resistivity and Seebeck coefficient of the thermoelectric (magnetic) material, respectively, and $r = (L_{\text{M}}^x / L_{\text{TE}}^x) \times (L_{\text{TE}}^y L_{\text{TE}}^z / L_{\text{M}}^y L_{\text{M}}^z)$ is the size ratio determined by the length of the thermoelectric (magnetic) material $L_{\text{TE(M)}}^{x,y,z}$ along the $x$, $y$, and $z$ directions. The second term on the right-hand side of Eq. (6) represents the STTG contribution and can reach the order of 100 $\mu$VK$^{-1}$ by optimizing the combination of the thermoelectric and magnetic materials as well as their dimensions. In fact, our experiments show that the Co$_2$MnGa/n(p)-type Si hybrid material exhibits a transverse thermopower of 82.3 $\mu$VK$^{-1}$ (−41.0 $\mu$VK$^{-1}$), which is one order of magnitude larger than the record-high $S_{\text{ANE}}$ value and is quantitatively consistent with the prediction based on Eq. (6) [Fig. 5(b)]. STTG appears in the absence of magnetic fields when the magnetic material layer has a finite coercive force and remanent magnetization, as demonstrated by the measurements using a $L1_0$-FePt/n-type Si hybrid material [Fig. 5(b)]. These results confirm the usefulness and potential of STTG. However, at present, such a large thermopower is obtained only when the combination of thermoelectric slabs and magnetic films, *i.e.*, the system with large *r* values, is used. For thermoelectric power generation (heat sensing) applications, it is necessary to realize large STTG in all-slab (all-film) hybrid materials with reasonable *r* values. Because STTG provides high flexibility for designing its performance based on a vast number of studies on the Seebeck effect and AHE, there is still plenty of scope for the improvement.

Finally, we would like to highlight an aspect that should be considered in ANE and STTG experiments. When these phenomena are applied in practice, it is preferable to utilize the synergistic effects of both. However, when the transport properties related to ANE and STTG are investigated quantitatively, they must be clearly and carefully separated using well-defined sample systems. The contamination of the STTG contribution in ANE experiments may occur accidentally, for example, when sample films are formed on doped Si substrates, which are often used in thermoelectrics and spin caloritronics. We verified that, even if thermally oxidized Si substrates are used, unintended electrical connections between thin films and doped Si may be formed through the side surfaces of the substrates, leading to significant errors in ANE results. Such an electrical connection can be avoided by patterning thin films using shadow masks or lithography techniques.

In this perspective, we have reviewed the progresses on the transverse thermoelectric generation using magnetic materials, with a particular focus on our recent activities. Systematic studies on ANE have clarified its mechanism and potential for next-generation thermoelectric technologies. By utilizing the emerging phenomena such as STTG, studies on transverse thermoelectric conversion will become more active in the near future; not only plain magnetic materials but also hybrid or composite materials will be key systems, where interface engineering is also important. For practical applications of the transverse thermoelectric generation, it is necessary to drive ANE and STTG in the absence of magnetic fields. Thus, searching and developing magnetic materials with large anomalous Nernst coefficients, large anomalous Hall angle, and high magnetic anisotropy are indispensable. The previous study suggests that magnetic materials with high magnetic anisotropy exhibit large ANE.[20] To understand the origin of this behavior and confirm its universality, one has to perform more systematic experiments using various materials, where a high-throughput screening method for ANE materials will be a powerful tool.[81] Although we have discussed ANE and STTG mainly in terms of the transverse thermopower, the reduction of the thermal conductivity of magnetic materials is also important for improving the efficiency of the transverse thermoelectric generation because the figure of merit for ANE and STTG is inversely proportional to the thermal conductivity in the same manner as the Seebeck effect.[62,75,77] To reduce the thermal conductivity, phonon engineering techniques and nanostructuring materials will be effective,[82,83] while they have rarely been used in spin caloritronics.[84] Therefore, for the further development of transverse thermoelectric generation technologies, the interdisciplinary fusion among spin caloritronics, nanoscale materials science, and thermal engineering is desired.

The authors thank R. Iguchi, K. Masuda, A. Miura, Y. Miura, and K. Yamamoto for valuable discussions and M. Isomura, N. Kojima, B. Masaoka, and R. Tateishi for technical supports. This work was partially supported by CREST "Creation of Innovative Core Technologies for Nano-enabled Thermal Management" (JPMJCR17I1) and PRESTO "Scientific Innovation for Energy Harvesting Technology" (JPMJPR17R5) from JST, Japan; Mitou challenge 2050 (P14004) from NEDO, Japan; Grant-in-Aid for Scientific Research (S) (18H05246) from JSPS KAKENHI, Japan; and the NEC Corporation.

## DATA AVAILABILITY

The data that support the findings of this study are available from the corresponding author upon reasonable request.

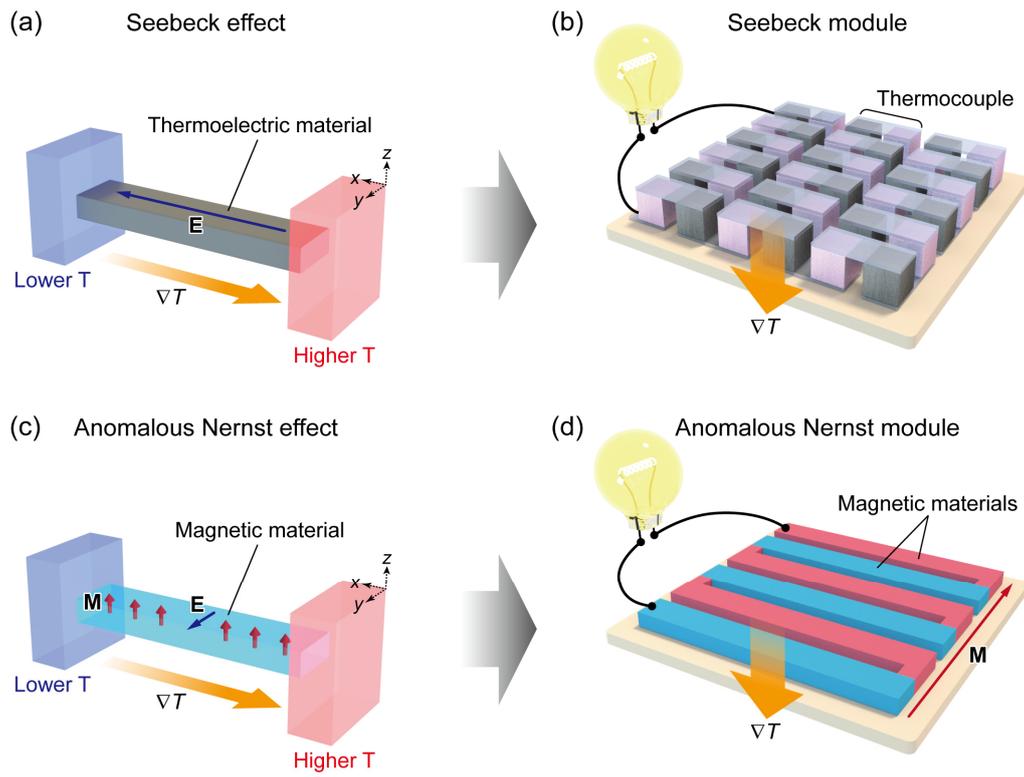

**FIG. 1.** Schematics of (a) the Seebeck effect, (b) the thermoelectric module based on the Seebeck effect, (c) ANE, and (d) the thermoelectric module based on ANE. **E**, ∇$T$, and **M** denote the electric field induced by the Seebeck effect or ANE, temperature gradient, and magnetization vector, respectively.

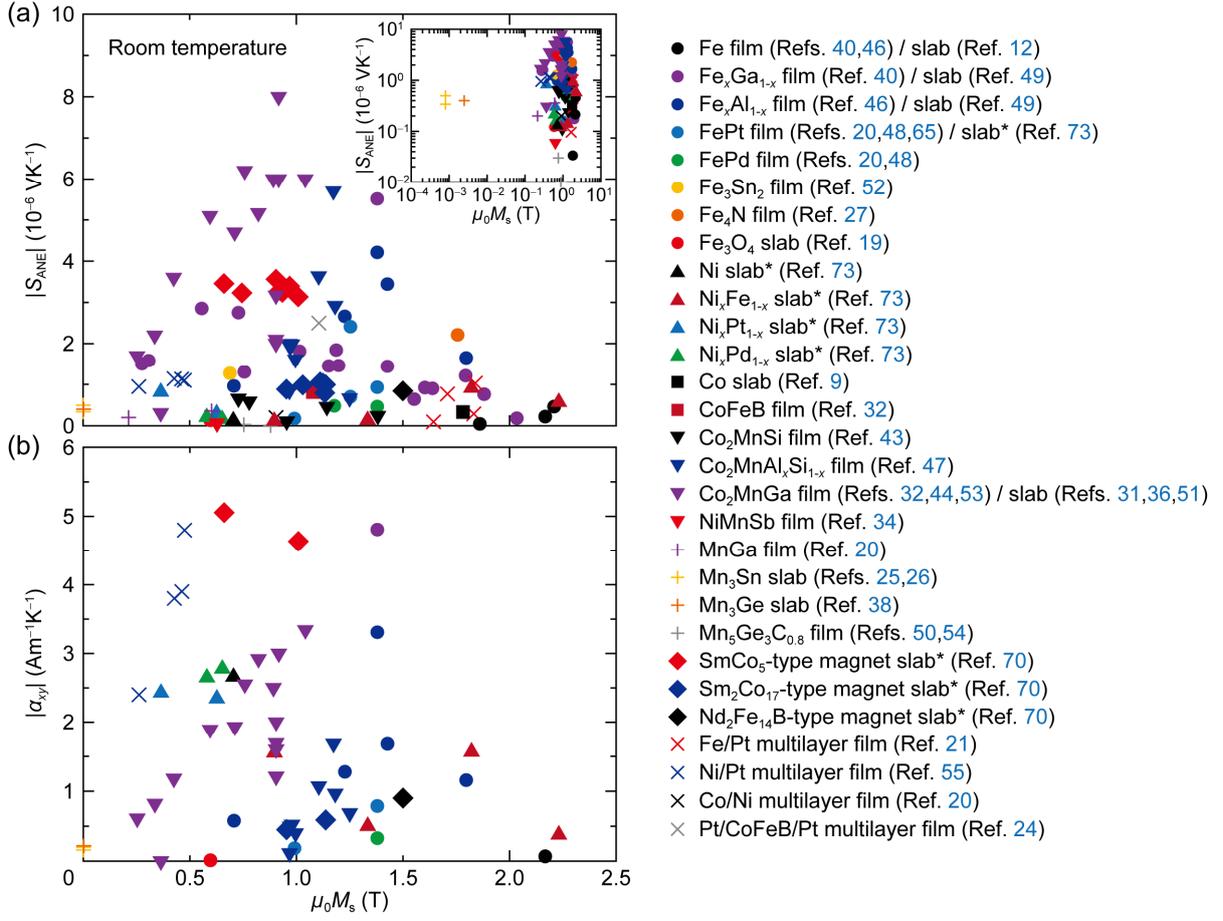

**FIG. 2.** Saturation magnetization $M_s$ dependence of the absolute values of (a) the anomalous Nernst coefficient $|S_{ANE}|$ and (b) the transverse thermoelectric conductivity $|\alpha_{xy}|$ for various magnetic materials. The inset of (a) shows the double-logarithmic $M_s$ dependence of $|S_{ANE}|$. $\mu_0$ is the vacuum permeability. $x$ in the legends denotes the composition ratio of the magnetic materials. Although ANE has been measured in a wide temperature range, only the data measured at room temperature are shown in this figure. The $|S_{ANE}|$ and $|\alpha_{xy}|$ values for the materials with an asterisk were estimated through the measurements of the anomalous Ettingshausen effect and the Onsager reciprocal relation. The $|\alpha_{xy}|$ value in Ref. 36 is corrected following the information in Ref. 51.

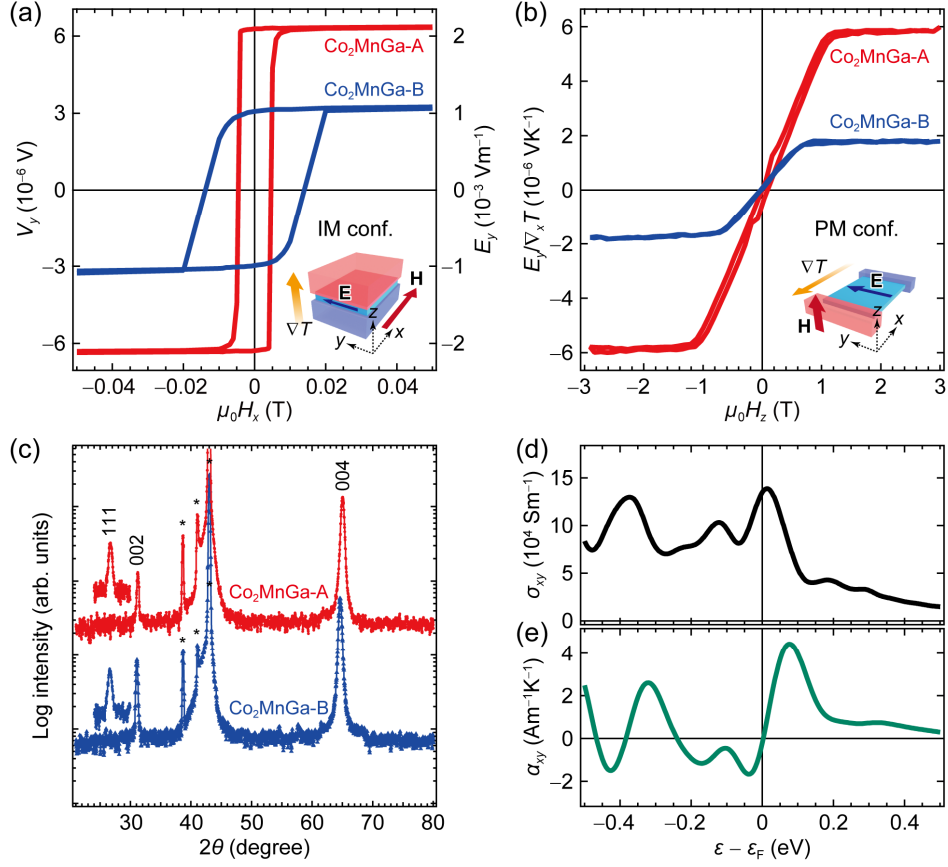

**FIG. 3.** (a) Magnetic field $\mu_0 H_x$ dependence of the transverse thermoelectric voltage $V_y$ and the transverse electric field $E_y$ in the Co$_2$MnGa-A and Co$_2$MnGa-B films at room temperature in the IM configuration, measured when the magnetic field and temperature gradient were applied along the $x$ and $z$ directions, respectively. A heater power of 160 mW was applied to the top heat bath to generate the temperature gradient along the $z$ direction $\nabla_z T$. (b) Magnetic field $\mu_0 H_z$ dependence of $E_y$ normalized by the temperature gradient $\nabla_x T$ in the Co$_2$MnGa-A and Co$_2$MnGa-B films at room temperature in the PM configuration, measured when the magnetic field and temperature gradient were applied along the $z$ and $x$ directions, respectively. $\nabla_x T$ was measured with an infrared camera by coating the sample surface with a black ink having high infrared emissivity. (c) XRD patterns for the out-of-plane direction ($\chi = 0°$) and the <111> direction ($\chi = 54.7°$) of the Co$_2$MnGa-A and Co$_2$MnGa-B films. The peaks denoted by asterisks arise from the diffractions from the MgO substrates. (d),(e) Intrinsic $\sigma_{xy}$ and $\alpha_{xy}$ for the stoichiometric Co$_2$MnGa as a function of the energy difference from the Fermi energy $\varepsilon - \varepsilon_F$, obtained from the first-principles calculations.[53]

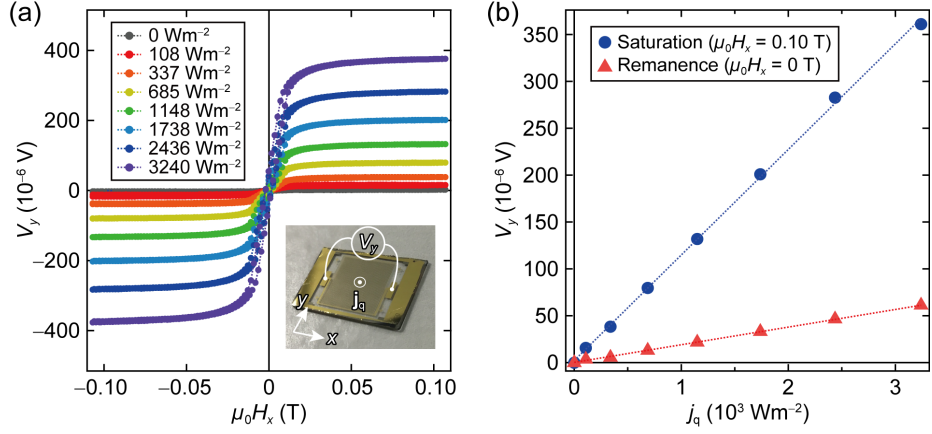

**FIG. 4.** (a) $\mu_0 H_x$ dependence of $V_y$ in the $Co_2MnGa$-Au thermopile device, in which 50 $Co_2MnGa$ wires are electrically connected in series through Au wires, for various values of $j_q$. $j_q$ denotes the magnitude of the heat flux density $\mathbf{j}_q$ along the $z$ direction. The $Co_2MnGa$ and Au wires align along the $y$ direction. (b) $j_q$ dependence of $V_y$ in the $Co_2MnGa$-Au thermopile device at the saturation magnetization ($\mu_0 H_x = 0.10$ T) and remanent magnetization ($\mu_0 H_x = 0$ T) states.

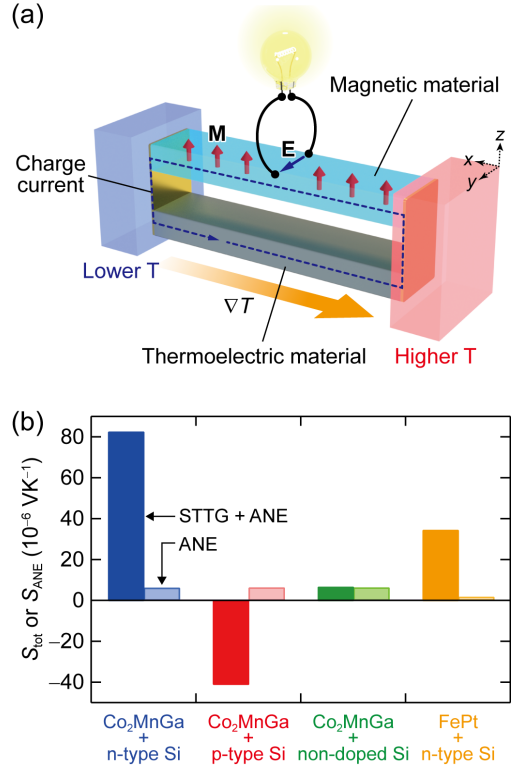

**FIG. 5.** (a) Schematic of STTG in a closed circuit comprising thermoelectric and magnetic materials electrically connected at both ends. (b) $S_{tot}$ or $S_{ANE}$ values at room temperature for the $Co_2MnGa$/n-type Si, $Co_2MnGa$/p-type Si, $Co_2MnGa$/non-doped Si, and FePt/n-type Si hybrid materials used in Ref. 75. When $S_{tot}$ ($S_{ANE}$) in the hybrid materials was measured, the thermoelectric and magnetic layers were electrically connected (disconnected). The $S_{tot}$ and $S_{ANE}$ values were estimated by extrapolating the magnetic field dependence of the transverse thermopower from the high field region to zero field.